\documentclass[12pt,a4paper]{article}
\usepackage{a4wide}
\usepackage{latexsym}
\usepackage{epsf}
\usepackage{amssymb}

\begin{document}

\def\pd{\partial}
\def\a{\alpha}
\def\b{\beta}
\def\g{\gamma}
\def\d{\delta}
\def\m{\mu}
\def\n{\nu}
\def\t{\tau}
\def\l{\lambda}

\def\s{\sigma}
\def\e{\epsilon}


\hyphenation{re-pa-ra-me-tri-za-tion}
\hyphenation{trans-for-ma-tions}


\begin{flushright}
IFT-UAM/CSIC-98-1\\
hep-th/9806075\\
June 11th 1998\\
\end{flushright}

\vspace{1cm}

\begin{center}

{\bf\Large String Representation of Wilson Loops}

\vspace{.5cm}

{\bf Enrique \'Alvarez}${}^{\diamondsuit,\clubsuit}$
\footnote{E-mail: {\tt ealvarez@daniel.ft.uam.es}},
{\bf C\'esar G\'omez}${}^{\diamondsuit,\spadesuit}$
\footnote{E-mail: {\tt iffgomez@roca.csic.es}} \\
\vspace{.3cm}
{\bf and  Tom\'as Ort\'{\i}n}${}^{\diamondsuit,\spadesuit}$
\footnote{E-mail: {\tt tomas@leonidas.imaff.csic.es}}

\vskip 0.4cm

${}^{\diamondsuit}$\ {\it Instituto de F\'{\i}sica Te\'orica, C-XVI,
  Universidad Aut\'onoma de Madrid \\
  E-28049-Madrid, Spain}\footnote{Unidad de Investigaci\'on Asociada
  al Centro de F\'{\i}sica Miguel Catal\'an (C.S.I.C.)}

\vskip 0.2cm

${}^{\clubsuit}$\ {\it Departamento de F\'{\i}sica Te\'orica, C-XI,
  Universidad Aut\'onoma de Madrid \\
  E-28049-Madrid, Spain}

\vskip 0.2cm

${}^{\spadesuit}$\ {\it I.M.A.F.F., C.S.I.C., Calle de Serrano 113\\ 
E-28006-Madrid, Spain}

\vskip 1cm


{\bf Abstract}

\end{center}

\begin{quote}
  
  We explore the consequences of imposing Polyakov's
  zig/zag-invariance in the search for a confining string. We first
  find that the requirement of zig/zag-invariance seems to be
  incompatible with spacetime supersymmetry. We then try to find
  zig/zag-invariant string backgrounds on which to implement the
  minimal-area prescription for the calculation of Wilson loops
  considering different possibilities.
  
\end{quote}


\newpage

\setcounter{page}{1}
\setcounter{footnote}{1}
\renewcommand{\theequation}{\thesection.\arabic{equation}}


\section{Introduction}

After the recent work of
Refs.~\cite{polyakov,gk,kn:RY,malda,gkp,witten,witten2,witten3,more}
we have for the first time a way to look for confining string
candidates by solving the loop equations of four-dimensional
non-Abelian Yang-Mills theories. An important ingredient that such
strings should satisfy is the so called zig/zag invariance: invariance
under generic reparametrizations included those not preserving
orientation of the parametrized loop \cite{polyakov}. Only after
fulfilling this symmetry requirement we can achieve the main goal of
satisfying the loop equations (see e.g.~\cite{migdal} and references
therein). In this paper we will discuss some of the implications of
zig/zag-invariance and its potential implementation in the framework
of the near-horizon geometry of D-brane solutions.



A longstanding problem in non-Abelian gauge theories is to find a string
representation of gauge non-local loop variables (Wilson loops),
satisfying the Polyakov-Makeenko-Migdal loop equations \cite{migdal}. 
It is known that fundamental strings, for instance Nambu-Goto strings,
are not suited for this purpose owing to the fact, recently stressed by
Polyakov \cite{polyakov}, of their failure to reproduce the
reparametrization invariance of the Wilson loop with respect to generic
reparametrizations, including those with vanishing derivative at some
point.

A Wilson loop is defined by

\begin{equation}
W(C) = {\rm Tr}\ U(C) = {\rm Tr}\ 
\left\{ P \exp{i\oint_{C} A_{\m}dx^{\m}}\right\} \, ,
\end{equation}

\noindent where the exponent is the integral of a one-form over
a closed path and, as such, it enjoys invariance under {\em all}
diffeomorphisms of the path, including those (orientation-reversing)
corresponding to reparametrizations $\xi\rightarrow \xi '$ where the
Jacobian changes sign.  This obviously implies:

\begin{equation}
W(C^{-1}) = W^{-1} (C)\, .
\end{equation}

Following Polyakov we will refer to this type of reparametrizations as
zig/zag symmetry. The net geometrical effect of this type of
reparametrizations is to induce foldings, hence a necessary condition
for a string theory to be a reasonable candidate of Yang-Mills string or
more generically of confining string, is to be blind to foldings
\cite{polyak}.

This is a very strong requirement on the open string sector of the
theory. In particular it means that all open string vertex operators,
which include factors of the square root of the determinant of the
worldsheet metric $h_{\alpha\beta}$ should decouple. This is equivalent
to say that small variations of the loop should be saturated, in the
corresponding string representation, by open string vertex operators
creating states in the first Virasoro level, which are the ones
representing the massless gluons. 
 
Those at higher levels, which represent massive states of higher spin,
are associated with vertex operators that must be integrated over the
boundary of the worldsheet using $\sqrt{h}$.

If some of this higher-level modes contribute to the variation of the
Wilson loop, then we necessarily lose zig/zag invariance (due to the
dependence of the vertex operator on the worldsheet metric) and very
likely we will find terms in the loop variation not consistent with the
Yang-Mills loop equations.

The simplest and more radical way to guarantee zig/zag invariance is,
of course, to require the string-induced metric on the boundary of any
open string worldsheet amplitude to vanish. In fact in this case any
vertex operator that potentially can spoil zig/zag invariance is
automatically decoupled from all possible open string amplitudes. A
question immediately arises: does the truncation imposed by
requiring zig/zag invariance, produce a consistent string theory?

We can now imagine a string metric background $G$ with ``horizons" to
be defined as subspaces of spacetime with vanishing pullback metric.
We could require these ``horizon" subspaces to contain time. In this
case the condition of zig/zag invariance for a string living in this
particular spacetime is tantamount to impose Dirichlet boundary
conditions on the coordinates transverse to the ``horizon". We refer
to these ``horizons'' as {\it zig/zag horizons}.

Of course, the string metric $G$ is not arbitrary and must satisfy the
generic constraints imposed by conformal invariance. So the first
problem we face in order to find a confining string is to discover a
consistent string metric, possessing a four dimensional spacetime
horizon to be identified with the spacetime where our Yang-Mills
theory is going to be defined.
\par
Following Polyakov we will identify the transversal coordinates with 
the Liouville field. The requirement of zig/zag invariance amounts to 
impose Dirichlet boundary conditions on the said Liouville coordinate. 
\par
It is important here to stress the difference between what we just
called zig/zag horizons and Dirichlet branes and near-horizon
geometries.

In a zig/zag horizon we impose Dirichlet boundary conditions on the
horizon in very much the same way as we do it for standard D-branes.
The main difference with D-branes is that now we impose the vanishing
of the string metric in the horizon in order to project out the open
string higher level modes, something that clearly we do not impose in
standard D-brane physics.

On the other hand, in the near-horizon approach to D-branes using
Maldacena's limit, massive modes are decoupled but the spacetime
picture of a D-brane is lost and replaced by Supergravity on a certain
$AdS_5\times S_5$ space-time.

It should be kept in mind that zig/zag horizons, being the locus for
Dirichlet boundary conditions, become in the same way as for generic
D-branes, a source for gravitons.  In fact it would be in principle
possible to compute, following Polchinski's prescription, some sort of
zig/zag-brane tension. Most likely zig/zag-branes could be simply
thought of as candidates for non-critical D-branes.


\section{Zig/Zag Invariance and Supersymmetry} 

In our previous discussion we simply addressed the question of zig/zag
invariance for the simplest bosonic case. A question that arises
immediately is the consistency of zig/zag invariance and
supersymmetry.

In the discussion of this issue it is important to separate worldsheet
supersymmetry and spacetime supersymmetry. The worldsheet supersymmetric
Lagrangian is given by

\begin{equation}
\begin{array}{rcl}
{\cal L}  & = & 
\sqrt{h}\left\{  h^{\a\b}\partial_{\a}X^{\mu}\partial_{\b}X_{\mu} 
+\frac{i}{2}\psi^{\m}\gamma^{\a}\nabla_{\a}\psi_{\m} 
\right. \\
& & \\
& & 
\left.
+\frac{i}{2}(\chi_{\a}\gamma^{\beta}\gamma^{\a}\psi^{\m})
(\partial_{\b}\chi_{\m} -\frac{i}{4}\chi_{\beta}\psi_{\m})
\right\}\, ,
\end{array}
\end{equation}

\noindent which implies that the zig/zag condition of tensionless open
strings does not clash with worldsheet supersymmetry. More precisely the
$\sqrt{h}$ factor, that spoils zig/zag invariance is common to both
the boson and fermionic worldsheet sectors. The only bosonic vertices
allowed by zig/zag invariance are the ones for massless vector bosons

\begin{equation}
\oint DX e^{i k\cdot X} d\xi d\theta\, ,
\end{equation}

\noindent with  $X$ being the string coordinate superfield and $D$ the 
worldsheet superderivative. In order to discuss spacetime supersymmetry
we should consider fermionic emission vertices. As it is well known the
definition of these vertices in the NSR formalism requires the use of
picture-changing manipulations \cite{kn:FMS}. In particular, for the
open string the fermion vertex of conformal dimension 1 is, in the
$-1/2$ picture, given by

\begin{equation}
\label{eq:fermionvertex}
\oint \sqrt{h}\ e^{-\frac{\phi}{2}} S^{\a} e^{i k\cdot X} d\xi\, ,
\end{equation}

\noindent where $\phi$ is the field introduced in the bosonization of
the ghost current and $S$ is the $SO(10)$ spinor. From the preceding
expression we observe that zig/zag invariance, at least naively, is
not consistent with spacetime supersymmetry, i.e.~with the existence
of non-vanishing open-string amplitudes for fermion vertex operators.
A way to see the potential difficulties for compatibility between
zig/zag invariance and spacetime supersymmetry is already suggested by
the asymmetric dependence on the worldsheet metric of the two terms
involved in the $\kappa$-invariant worldsheet action, where the extra
Wess-Zumino term is manifestly independent on the worldsheet metric
and, therefore, zig/zag invariant.\footnote{In the light-cone gauge
  the fermionic vertex operator is like the one in equation
  (\ref{eq:fermionvertex}) without the spinor ghost and with an
  $SO(8)$ spinor field that behaves as a $1/2$ spinor in two
  dimensions, which makes clear the dependence on the determinant of
  the metric in the integrated version}

A different issue we can worry about is the fate of tachyons in
zig/zag-invariant strings.  It is interesting to observe that both in
the standard bosonic case and in the NSR superstring, before imposing
GSO projection, the tachyon vertex operators are not consistent with
zig/zag invariance, since in both cases they depend on the determinant
of the worldsheet metric $h_{\alpha\beta}$

\begin{equation}
\left\{
\begin{array}{c}
\left.\oint \sqrt{h}\ e^{ik\cdot X}\  \right|_{k^2=-2}\, , \\
\\
\left. \oint \sqrt{h}\ k\cdot\psi e^{i k\cdot X}\ \right|_{k^2 =-1}\, ,\\
\end{array}
\right.
\end{equation}

\noindent and therefore they should decouple by the zig/zag mechanism.

In summary, what we conclude from our previous discussion is that a
confining string satisfying zig/zag invariance, even if it is
world-sheet supersymmetric, is truncating the open string spectrum to
pure gluons only.

As mentioned above, we can try to understand this phenomenon from the
point of view of $\kappa$-symmetry. In fact, space-time supersymmetry
amounts to consider loops in superspace and to work directly with the
Green-Schwarz superstring.  Imposing zig/zag invariance in the
Green-Schwarz superstring is equivalent to reduce ourselves to the
topological Wess-Zumino term in what concerns the open string sector,
which in turn implies that we lose $\kappa$-invariance.


\section{Confining Strings and Near Horizon Geometry}

Polyakov's Ansatz for confining strings is based on a non-critical
string in four dimensions, supplemented by a Liouville field $\varphi$
so the the effective 5-dimensional metric has the form \cite{polyakov}

\begin{equation} 
\label{eq:ansatz}
ds^2 = a^{2}(\varphi) dx_{\parallel}^2 -d\varphi^2\, ,
\end{equation} 

\noindent satisfying the horizon condition:

\begin{equation} 
a(\varphi_{0}) = 0\, .
\end{equation} 

Zig/zag symmetry is implemented by requiring the gauge fields to live
on the horizon. This is equivalent to impose the following Ansatz for
the Wilson loop:

\begin{equation} 
W(C)  = 
\int
\mathcal{D}X \mathcal{D}\varphi\
\exp{\left\{ -\int_{\Sigma} G_{\m\n}(X,\varphi)
\partial X^{\m}\bar{\partial}X^{\n}
+T(X,\varphi)\partial \varphi \bar{\partial}\varphi 
+\ldots\right\} }\, .
\end{equation} 

\noindent where one integrates over all embeddings $X^{\mu}$
of the worldsheet whose boundary is on the horizon
$\varphi=\varphi_{0}$ and is the loop $C$. In this expression the part
of the metric Eq.~(\ref{eq:ansatz}) parallel to the brane (the
4-dimensional spacetime where we want to calculate Wilson loops),
$a^2 dx_{\parallel}^2$ has been rewritten as $G_{\m\n} dX^{\m}
dX^{\n}$).  The consistency conditions on the metric will be the ones
implied by conformal invariance of the closed string sector.


A very different computation of the Wilson loop based on confining
strings, inspired on the holographic map and on previous work on
absorption coefficients for non-dilatonic D-branes \cite{gk} was
developed by Rey, Yee and Maldacena \cite{kn:RY,malda} and further
extended by Witten \cite{witten}.  In that approach the spacetime
metric that enters the string sigma model is the one that describes
the near-horizon geometry of the D-3-brane: $AdS_{5}\times S^{5}$,
which is assumed to lead to a conformally-invariant theory. The 
$AdS_{5}$ part 

\begin{equation} 
ds^2 = \frac{R^2}{z^2}\left(dx_{\parallel}^2 -dz^2\right) 
-R^2 d\Omega_5^2\, ,
\end{equation} 

\noindent plays the most important role. It can be written in the 
from of Eq.~(\ref{eq:ansatz}) with the change of variables

\begin{equation} 
\label{eq:eso}
z = R e^{\varphi/R}\, .
\end{equation} 

\noindent but the fifth coordinate does not have a Liouville field 
interpretation. Furthermore, they proposed to place the loop $C$ in
the {\em far from the brane region} of $AdS$ spacetime (that is: far
from the zig/zag horizon).  A semiclassical approximation to the
Wilson loop is given by the minimal area surface in $AdS_{5}$ whose
boundary is the loop $C$.  Renormalization is necessary and it
corresponds to the subtraction of the length of the loop in the
$AdS_{5}$ metric at the far-from-the-brane boundary region.

This definition of the Wilson loop is manifestly not zig/zag-invariant
since the pullback metric at the boundary space where the loop is
located is not vanishing. Notice that the problem with
zig/zag-invariance is not only due to the modified Wilson loop, used
in \cite{kn:RY,malda}, that includes the quantum numbers associated
with the transversal coordinates in $S^5$, but is a problem even for
the finite-temperature models where supersymmetry is explicitly broken
and effective decoupling of Higgs fields is assumed, as it is done in
\cite{witten,more}.

We can now compute the semiclassical approximation to the
zig/zag-invariant Wilson loop for the $AdS_{5}$ confining string. For
simplicity we will consider a rectangular Wilson loop in spacetime
with $\ell$ being the separation between the static {\em quarks}
(actually, infinitely heavy gauge bosons).  The zig/zag-invariant
Wilson loop is then given on the horizon by

\begin{equation}
\label{eq:embedding}
\left\{
\begin{array}{rcl}
\tau & = & X^{0}\, , \\
& & \\  
\sigma & = & X\, ,\\
\end{array}
\right.
\end{equation}

\noindent where $X$ is one of the spacelike coordinates and 
$\sigma\in [-\ell/2,+\ell/2]\, ,\tau\in[0,T]$.

The semiclassical approximation will be determined by the minimal area
surface corresponding to the profile $ \varphi = \varphi(\sigma)$
(with the boundary conditions $\varphi(\pm \ell/2) = \varphi_{0}$.
The induced metric on the worldsheet is given by

\begin{equation}
ds^2 = a^{2}d\tau^2 - \left(a^{2} - \varphi_\sigma^2\right)d\sigma^2\, .
\end{equation}

\noindent where $\varphi_\sigma \equiv d\varphi/d\sigma$.

The corresponding Nambu-Goto action will then be the integral of the
determinant of the induced metric, that is:

\begin{equation}
S = T \int_{-\frac{\ell}{2}}^{\frac{\ell}{2}} 
d\sigma\ a \sqrt{a^2 +\varphi_\sigma^2}\, .
\end{equation}

The {\em energy} associated to translations in $x$ is given by:

\begin{equation}
E = - \frac{a^3}{\sqrt{a^2    + \varphi_\sigma^2}}\, .
\end{equation}

The most general solution to this problem is a string ``ironed'' over
the horizon:

\begin{equation} 
a = 0\, .
\end{equation} 

It is easy to prove that this result is general for any Polyakov
confining string Ansatz, because from energy conservation we easily get:

\begin{equation}
\varphi_\sigma^2 = a^2 \left(\frac{a^4}{E^2} - 1\right)\, ,
\end{equation}

\noindent  constraining the motion to lie on the region $a^2 > E$. 
The only solution which can be attached to the horizon at the
endpoints $\sigma = \pm \ell/2$ is the trivial one $a = 0$.

\section{Exploring other Possibilities}

In this section we will explore other ways to implement this set of ideas
different from using $AdS_{5}$ as a string background (supplemented by an
$S_{5}$ ``hidden sector'').

The first thing one can try is to find directly solutions of the
$\beta$-functions (with no hidden sector) equations satisfying all the
requirements. Thus, we can look for solutions of the equations of motion
corresponding to the action

\begin{equation}
S= \int d^{d}x\sqrt{|g|}\ e^{-2\phi}
\left[ R -4\left(\partial\phi\right)^{2}
+{\textstyle\frac{1}{2\cdot 3!}}H^{2} 
+{\textstyle\frac{(d-26)}{3\alpha^{\prime}}}
\right]\, ,
\end{equation}

\noindent of the form (\ref{eq:ansatz}). These solutions necessarily
have a non-trivial dilaton field, due to the dilaton potential
proportional to  the  central-charge deficit. We will not consider
non-trivial Kalb-Ramond fields. The simplest solutions are, then,
domain-wall-type solutions. These have been intensively studied in
Refs.~\cite{kn:CLPSST} for the generic model

\begin{equation}
S= \int d^{d}x\sqrt{|g_{E}|}\
\left[ 
R_{E} +{\textstyle\frac{1}{2}}
\left(\partial\varphi\right)^{2}
+{\textstyle\frac{1}{2\cdot 3!}}H^{2} 
+{\textstyle\frac{1}{2}}m^{2}e^{-a\varphi}
\right]\, .
\end{equation}

Real solutions exist in any dimension for strictly  negative values of
the parameter

\begin{equation}
\Delta=a^{2}-2{\textstyle\frac{(d-1)}{(d-2)}}\, .
\end{equation}

If we reverse the sign of the scalar potential (or, equivalently we make
$m$ imaginary), then, solutions exist for any strictly positive value of
$\Delta$. If we rescale our action to the Einstein frame and then we
canonically normalize the dilaton field, as above,, we find that our
case is mixed, and we can only find real solutions if we use as
coordinate transverse to the domain wall a {\it timelike} coordinate. If
$x$ is such a coordinate, then, the solution looks in the Einstein frame

\begin{equation}
\left\{
\begin{array}{rcl}
ds^{2}_{E} & = & H^{2} \left[ dt^{2} -d\vec{y}^{\ 2}_{d-2}\right] 
+dx^{2}\, ,\\
& & \\
e^{\phi} & = & H^{-\frac{(d-2)}{2}}\, ,\\
\end{array}
\right.
\end{equation}

\noindent where $H=bx+c$, where the constant $b$ is a function of the
dimension, $m$ and the parameter $a$ and $c$ can be set to zero by a
coordinate shift. Rescaling the solution back to the string frame and
after a simple coordinate change we find 

\begin{equation}
\left\{
\begin{array}{rcl}
ds^{2} & = & dt^{2} -d\vec{y}^{\ 2}_{d-2} +dz^{2}\, ,\\
& & \\
e^{\phi} & = & e^{-\frac{(d-2)}{2}H}\, .\\
\end{array}
\right.
\end{equation}

This solution is similar to the D-instanton solution in the sense that
both have a flat spacetime. Precisely for this reason, it is totally
inappropriate for our purposes.

With our Ansatz, the above solution is unique. Considering a non-trivial
2-form field would privilege a 2-dimensional subspace and this would
not be consistent with the Ansatz. Thus, we have to look for other kinds
of solutions, now with an appropriate hidden sector that makes zero the
total central charge.

First, it is easy to see, that there are no vacuum solutions (apart from
Minkowski) of the string $\beta$-functions of the form
(\ref{eq:ansatz}). Thus, it is natural to explore other possibilities
relaxing some of the conditions that the string background has to
satisfy.

If we relax the condition that the function $a$ that vanishes on the
horizon is a common factor for the metric of the whole 4-dimensional
space, we find a vacuum solution which is unique and depends on only
one integration constant:

\begin{equation}
\label{eq:vacuumsolution}
ds^{2} = a^{2}dt^{2} -d\vec{y}^{\ 2}_{3} \pm d\varphi^{2}\, ,
\hspace{1cm}
a= 1+b\varphi\, .
\end{equation}






After a straightforward calculation we find that the action is, up to
a numerical coefficient

\begin{equation}
S \sim  T b \ell^{2}\, .
\end{equation}

This background solves the $\beta$-functions and leads to the area law
but it does not respect zig/zag-invariance. This result is independent
of the dimension (unless the constant $b$ contains information about
it in a way unknown to us).

The next possibility is to add a non-trivial dilaton. Using Polyakov's
Ansatz (\ref{eq:ansatz}) we find in $d$-dimensions a unique solution

\begin{equation}
ds^{2} = a^{2\alpha}\left(dt^{2} -d\vec{y}^{\ 2}_{3}\right) 
\pm d\varphi^{2}\, ,
\hspace{1cm}
e^{\phi-\phi_{0}} = a^{\beta}\, ,
\end{equation}

\noindent where $a$ has the same form as in the previous 
solution and $\alpha,\beta$ are two numerical constants that depend on
the dimension:

\begin{equation}
\left\{
\begin{array}{rcl}
\alpha & = &  1- \frac{(d-2)}{(d-1)}
\frac{1}{1\pm \sqrt{\frac{2}{(d-2)(d-1)}}}\, , \\
& & \\
\beta & = & \pm \sqrt{\frac{2(d-2)}{(d-1)}} 
\left[\frac{1}{(d-1)} \pm \sqrt{\frac{2}{(d-2)(d-1)}} \right]\, .\\
\end{array}
\right.
\end{equation}

\noindent (The two possible signs in these constants are unrelated to the
two possible signatures).

As we explained in the Introduction, there is no solution for
the string action in this background.

Again, we have to relax the conditions satisfied by our metric. There is 
a whole family of solutions that we can now use but we will focus in a
particular one: the T~dual of the vacuum solution
Eq.~(\ref{eq:vacuumsolution}) in the time direction:

\begin{equation}
\label{eq:tdualvacuumsolution}
ds^{2} = a^{-2}dt^{2} -d\vec{y}^{\ 2}_{3} \pm d\varphi^{2}\, ,
\hspace{1cm}
e^{\phi-\phi_{0}} = a\, .
\end{equation}

Now there is solution only for the lower sign:

\begin{equation}
a=b\sqrt{\left(\ell/2\right)^{2}-\varphi^{2}}\, ,
\hspace{1cm}
S\sim \frac{1}{b^{3}\ell^{2}}\, ,
\end{equation}

\noindent which doesn't give the are law.

We would like to stress two facts with respect to this solution:

\begin{enumerate}

\item The behavior is completely different from that of its T~dual. This
may be yet another example of T~duality not working in the time
direction.

\item The sign of the dilaton field can be reversed $\phi^{\prime}$. 
This allows us to fit the Yang-Mills $\beta$-function:

\begin{equation}
\beta=\frac{\partial g}{\partial \log{\mu}}
= \frac{\partial e^{\phi^{\prime}/2}}{\partial\varphi}\sim a^{-3/2}\sim
e^{3\phi^{\prime}/2}\sim g^{3}\, .
\end{equation}

\end{enumerate}

\section{Liouville Dressing}

A different attempt to find a confining solution enjoying
zig/zag-invariance would imply a modification, by introducing a
Liouville potential, of the worldsheet Lagrangian on the near-horizon
geometry for D-3-branes\footnote{In the Liouville framework we are
  forced to consider ``non-critical'' D-branes with the Liouville
  field in the transverse direction. It would be interesting to work
  out, for these non-critical D-branes, the effect of tadpoles for
  macroscopic string states \cite{kn:seiberg}.}.

The logic behind this modification is as follows. In the case of
D-3-branes the geometry near the horizon is $AdS_5\times S^5$ with the
$S^5$ parametrizing the extra scalars required for $N=4$ supersymmetry.

On the other hand, in the original Polyakov approach to confining
strings, a non-critical string with four spacetime coordinates and one
extra Liouville coordinate is used. This Liouville coordinate is
constrained by the condition of vanishing $\beta$-functions, which, in
principle, can be implemented by using some string effective action, and
by the requirement of vanishing central extension.

A potential zig/zag interpretation of the $AdS_5\times S^5$ near-horizon
metric amounts to identify the radial coordinate with the Liouville
field as done in Eq.~(\ref{eq:eso}).

However in order to make more precise this identification we should
correct the Liouville field in a way that effectively mimics the
contribution to the central extension of the extra coordinates
parametrizing the $S^{5}$. The most naive way to achieve that is to
use a Liouville field with central extension $c_L = 6$. In the
conformal gauge such a Liouville field is defined by

\begin{equation}
S_{\rm Liouville}\equiv \frac{1}{4\pi\gamma}
\int d\sigma d\tau\left\{
{\textstyle\frac{1}{2}}[\left(\partial_{\tau}\varphi \right)^{2}
-\left(\partial_{\sigma}\varphi \right)^{2}]
-\mu^2 e^{2\varphi}\right\}\, .
\end{equation}

\noindent The classical central extension is given by $c=3/\gamma$.
If we rewrite $c=6$ we get $\gamma=1/2$. If, instead, we use the
quantum central extension $c=3+ 1/\gamma$ then we must set $\gamma =
1/3$.  This difference will not be relevant for our qualitative
discussion of the Wilson loop and we will use the classical value
$\gamma=1/2$ only.  The factor $\mu^{2}$ in the above action is a
worldsheet cosmological constant term. As we are going to see in a
moment, non-trivial solutions for zig/zag-invariant Wilson loops
require the cosmological constant to be negative (as above).

In a certain sense the addition of the Liouville potential amounts to
include a sort of gravitational dressing to the Wilson loop,
preserving zig/zag-invariance. The roughest approximation we can think
of for the computation of Wilson loops for static quarks will then
consist in taking the $AdS$ metric and simply adding the Liouville
potential

\begin{equation}
S = \frac{1}{4\pi}\int  d\sigma d\tau 
\left[ a^2 (\phi)(\partial x_{\parallel})^2 \right]
+S_{\rm Liouville}\, ,
\end{equation}

\noindent where 

\begin{equation}
a(\varphi) = e^{- \varphi/R}\, .  
\end{equation}

Observe that we are identifying the $\varphi$ spacetime coordinate
with the Liouville field and the natural scale for the spacetime
coordinate is the $AdS$ radius $R$ so that is also the natural scale
for the Liouville. In what follows we will work in units in which $R=1$.

Notice also that we are working in the conformal gauge where the
Liouville mode is precisely the conformal factor of the worldsheet
metric, so that instead of minimizing the Nambu-Goto action plug our
Ansatz into the above action and look for solutions of the resulting
equations of motion. There is also a first integral given
by:

\begin{equation}
E = -\frac{1}2{\gamma}(\partial_{\sigma}\phi)^2 
-2 a^2 +\frac{\mu^2}{\gamma}e^{2\phi}
\end{equation}

The solution satisfying the zig/zag constraints is given in terms of
the following elliptic integral

\begin{equation}
x - \ell/2 = - \frac{1}{\mu\sqrt{2}\sqrt{t_{-}^2 + t_{+}^2}}\
{\bf ds}^{-1}\left(\left.\frac{e^{\phi}}{ \sqrt{t_{-}^2 
+t_{+}^2}}\,\right| \frac{t_{-}^2}{ t_{-}^2 + t_{+}^2}\right)\, ,
\end{equation}

\noindent where ${\bf ds}$ is the Jacobi elliptic function which has a
zero at the point $K+iK^{\prime}$ (where $K$ and $K^{\prime}$ are
respectively the real and imaginary quarter periods) and a pole at the
origin.  $t_{+}^{2}$ and $-t_{-}^{2}$ are the solutions of the
equation

\begin{equation}
t^{4} -\frac{\gamma E}{\mu^2} t^{2} - \frac{2\gamma}{\mu^2} = 0\, .  
\end{equation}

Giving the fact that $ 4 K$ is a real period of the Jacobi function,
where $K$ is given in terms of the {\em modulus} $
m\equiv\frac{t_{-}^2}{ t_{-}^2 + t_{+}^2}$ by

\begin{equation}
K(m) \equiv \int_{0}^{\pi/2} \frac{d\theta}{\sqrt{1 - m^2 \sin^2{\theta}}}\,
\end{equation}

\noindent we can enforce another zero at $x= -\ell/2$ by equating

\begin{equation}
4 \frac{K(m)}{\mu\sqrt{2}\sqrt{t_{-}^2 + t_{+}^2}} = \ell\,
\end{equation}

\noindent which gives the {\em energy} $E$ as a function of $\ell$.

In the limit $E\rightarrow\infty$ $t_{+}\sim \gamma E$ and $t_{-}\sim
0$. This limit corresponds to $\ell = \frac{2\pi}{\sqrt{2\gamma E}}$,
up to the point that all our approximation can not be trusted anymore.
\par
Using the fact that the Jacobi function in this limit (for zero
modulus) reduces to $\frac{1}{\sin}$, we get for the action the
following result

\begin{equation}
W(C) = \frac{-2\pi^2 T}{\gamma\ell}\, ,
\end{equation}

\noindent If we were to cut off the integral over 
$\varphi$ at $\varphi=0$ nothing would change much. This can be easily
seen to be equivalent to a maximum allowed value for the energy $E$.

This is a non-confining solution that very likely reflects the fact
that we are working with $AdS$ spacetime. The effect of the Liouville
mode simply reduces to allow a zig/zag invariant solution. 

Physically, thus, the Liouville term with negative cosmological
constant acts as a sort of ``potential well'' allowing the existence
of ``bound states'' in a mechanism somewhat similar to the one
proposed by Witten in his black-hole-type solutions \cite{witten2}.

It is worth noticing that the potential problems concerning quantum
Liouville theory would be important for the closed string sector in
the bulk.  For the open string sector we can expect the effect of
Liouville tachyons to be suppressed on the horizon as a consequence of
zig/zag-invariance (the metric on the horizon goes to zero as the
inverse of the Liouville potential).


\section{Conclusion}

The aim of this paper has been to investigate Polyakov's approach to
confining string representations of Wilson loops, based on imposing
zig/zag invariance. Next we summarize the main conclusions of our
research.  

First of all we have observed, by working out several examples, that
the conditions of zig/zag-invariance impose very severe constraints on
the spacetime metrics. In fact, without modifications, either
involving a non-trivial dilaton background, or some type of Liouville
dressing, we have been unable to find a satisfactory answer.  This
situation contrasts with the potentially easy task of finding good
string representations for Wilson loops, even in the case in which
supersymmetry is broken by the mechanism of Witten \cite{witten2},
once the condition of zig/zag invariance is relaxed.

Thus the whole problem boils down to understand the physical relevance
of such strong reparametrization invariance requirement. It seems
clear that only zig/zag-invariant Wilson loops have a chance to
satisfy Makeenko-Migdal-Polyakov loop equations. If this condition is
implemented by forcing the Yang-Mills fields to live on some horizon
hypersurface,then we are forced to modify drastically the
holographic map, if we want to make it consistent with generic zig/zag
reparametrizations. Geometrically, in the context of near-horizon
D-brane geometries, this turns out to be equivalent to define the
four-dimensional theory on the horizon and not on the boundary as it
is compulsory in the holographic approach.  

Zig/zag-invariance is, however, an ingredient that strongly affects the
definition of a consistent string theory. In particular, if we reduce
ourselves to the bosonic string the condition of zig/zag changes the
way we should interpret the tachyon instability. In fact if we succeed
in obtaining a bosonic zig/zag invariant string the tachyon vertex
would not appear since it is not zig/zag-invariant. In other words,
this general-reparametrization invariance acts as a projector that
automatically suppress the tachyon from the bosonic string spectrum.

This issue can be potentially interesting since the critical dimension
$d=26$ of the bosonic string is already magically attached to the
coefficient of the pure Yang-Mills $\beta$-function in $N=0$. Namely
the one-loop Yang-Mills $\beta$-function \cite{kn:MV}

\begin{equation}
\beta = -\frac{g^{3}}{16\pi^{2}}\frac{C_{2}(G)}{6}
\left[22 -4\nu(M)-\nu(R)\right]\, ,
\end{equation}

\noindent  (where $\nu$ is the number of Majorana fermions ($M$) 
or real scalar fields ($R$)) vanishes in the case in which there are
22 scalar and no fermion fields.  which are precisely the number of
transversal coordinates for a (bosonic) D-3-brane in 26 dimensions.

\noindent

The reason nobody cares about this purely $N=0$ theory defined on the
world volume of a bosonic D-3-brane is of course because of the
tachyon instability. It is precisely at that point where the relevance
of zig/zag symmetry as a potential way to project out the tachyon, for
the open string sector, may become relevant.

The other question we have observed is that zig/zag-invariance seems to
differentiate between the worldsheet supersymmetry and the spacetime
supersymmetry. As discussed above this is due to the form of fermionic
vertex operators. Assuming that for a moment it seems that the
requirement of zig/zag invariance not only projects out the open
string tachyons but also ``breaks'' spacetime supersymmetry.  This
``breaking'' can be considered dynamical only if we have some
dynamical procedure to implement the zig/zag invariance. In this paper
we have suggested a wild and, needless to say, very primitive
possibility, based on the observation that the metric on the world
volume for the supergravity solutions for D-branes is zero, and in
that way a candidate to zig/zag invariant metric. 

Finally, a last comment on our Liouville approach and $N=0$ solutions.
In order to go to $N=0$ the most natural approach is, following
Witten's suggestion, to decouple the extra matter in the $N=4$
supermultiplets. The Liouville attempt consists in replacing the
geometry used in performing this decoupling by simply introducing a
Liouville mode contributing in the appropriated way to the central
charge. We can even imagine that once we move in the Liouville
direction (actually going from the horizon to the boundary) we reach a
region where the Liouville mode is effectively decoupled,
interpolating in a certain way between the zig/zag invariant horizon
region with $N=0$ and a boundary region with conformal invariance.


\section*{Acknowledgments}

The work of E.A., C.G.~and T.O.~has been partially supported by the
European Union TMR program FMRX-CT96-0012 {\sl Integrability,
  Non-perturbative Effects, and Symmetry in Quantum Field Theory} and
by the Spanish grant AEN96-1655.  The work of E.A.~has also been
supported by the Spanish AEN96-1664.


\appendix


\section{Global Structure of $AdS$ and $CAdS$}

Given the basic importance of Anti de Sitter metric in the whole 
description of the spacetime region close to the brane,
 we have collected here some geometric facts, relevant for the 
discussion of boundary conditions in the main text.

Anti de Sitter space in p dimensions ($AdS_p$) is defined
(\cite{gibbons}) as the induced metric on the hyperboloid $(X^0)^2 +
(X^p)^2 - \d_{ij} X^i X^j = 1; (i,j = 1\ldots p-1)$ embedded in an
ambient space $\mathbb{R}_{2,p-1}$ (that is, $\mathbb{R}_{p+1}$
endowed with a Minkowskian metric with two times, $ds^2 = (dX^0 )^2 +
(dX^p )^2 -\d_{ij} dX^i dX^j $. Defined in that way, it clearly has
topology $S^1\times \mathbb{R}^{p-1}$ (as well as closed timelike
curves).  The universal covering space ($CAdS_p$) has topology
$\mathbb{R}^p$.

This definition makes it manifest the underlying $O(2,p-1)$ symmetry.
As it is well known, there is a $2-1$ correspondence between
$O(2,p-1)$ and the conformal group of Minkowski space in $p-1$
dimensions, $C(1,p-2)$.

The {\em boundary} at infinity $\partial AdS$ can be defined as the
region where all $X^{\m}$ are rescaled by an infinite amount,
$X^{\m}\rightarrow \xi X^{\m}$, where $\xi\rightarrow \infty$ .
In that way, the boundary is characterized by the relationship 
$(X^0)^2 + (X^p)^2 - \d_{ij} X^i X^j = 0$, which is nothing but the well-known
$O(2,p-1)$ null-cone compactification of Minkowski space, $M^{\mathbb{C}}$.
\cite{penrose}.The way it works 
is that to any regular point of Minkowski space,$x^{\m}\in M$, there
corresponds another point in $M^{\mathbb{C}}$, namely 

\begin{equation}
\left\{  
\begin{array}{rcl}
X^0 &=& x^0\, ,\\
& & \\
X^i &=& x^i\, ,\\
& & \\
X^{p-1} &=& \frac{1+ x^2}{2}\, ,\\
& & \\
X^p &=& \frac{1 - x^2}{2}\, .\\
\end{array}
\right.
\end{equation}

The points in $M^{\mathbb{C}}$ which are not in $M$ correspond to $X^p
+ X^{p-1} = 0$. This means that this compactification amounts to add
an extra null cone at infinity.

The $AdS$ metric can be easily put in the {\em globally static form} by means of
the ansatz

\begin{equation}
\left\{
\begin{array}{rcl}
X^0 &=& \cos \t \cosh \chi\, ,\\
& & \\
X^p &=&\sin \t \cosh \chi\, ,\\
& & \\
X^i &=& \sinh \chi n^i\, ,\\
\end{array}
\right.
\end{equation}

\noindent where $\d_{ij}n^i n^j = 1, (i,j = 1,\ldots,p-1)$. The result is

\begin{equation}
ds^2 = (\cosh \chi)^2 d\t^2 - (d\chi)^2 - (\sinh \chi)^2 d\Omega_{p-2}^2\, .
\end{equation}

\noindent $AdS$ corresponds to $0 \leq \t \leq 2\pi$, and $CAdS$ to 
$0 \leq \t \leq \infty$

The antipodal map $ J: X\rightarrow - X$, corresponds in this
coordinates simply to $ (\t,\chi,\vec{n}) \rightarrow (\t + \pi,\chi,
- \vec{n})$.

A different , but closely related set of coordinates are the ones used
by Susskind and Witten \cite{gibbons}. The metric has the form

\begin{equation}
ds^2 = \frac{R^2}{(1 - r^2)^2}( 4 \sum_{i=1}^{p-1}
(dx^i)^2 - (1 + r^2)^2 dt^2)\, .
\end{equation}

They are easily obtained from the globally static form by

\begin{equation}
\sinh{\chi} = \frac{2r}{1- r^2}\, .
\end{equation}

$CAdS$ itself corresponds to the ball $r < 1$, and the boundary sits on
the sphere $r = 1$.

Another interesting set of coordinates (common to all constant curvature spaces)
is Riemann's, in which the metric reads

\begin{equation}
ds^2 = \frac {\eta_{\m\n}dy^{\m}dy^{\n}}{(1 - \frac{r^2}{4 R^2})^2}\, ,
\end{equation}

\noindent where $\m ,\n, 0,\ldots p-1$, $\eta_{\m\n}$ is the ordinary Minkowski metric,
and $r^2 \equiv \eta_{\m\n}y^{\m}y^{\n}$.

This coordinates are quite natural in the following sense. They can be constructed
(\cite{ss}) by 

\begin{equation}
y^{\m} \equiv 2 R u^{\m} \tanh \frac{s}{2R}\, ,
\end{equation}

\noindent where $u^{\m}$ is the unit tangent vector to the geodesic going 
to the point P from a fiducial point $P_0$; and $s$ is the geodesic
distance from $P_0$ to $P$, and using the fact that the geodesic
deviation between neighboring geodesics grows as $\eta = R
(\frac{d\e}{ds})_{s=0} |\sinh \frac{s}{R}|$, and that the angle
between the tangents to such neighboring geodesics is precisely the
volume element on the unit Minkowskian sphere, which can be easily
obtained in terms of the ordinary volume on the unit Euclidean sphere:
$d\Omega^2_{p-1}(hyperbolic)\equiv - d \xi^2 - \sinh^2 \xi
d\Omega^2_{p-2}$.

In that way

\begin{equation}
ds^2 = dr^2 - R^2 \sinh^2 \xi \frac{r}{R} (d \xi^2 + \sinh^2 \xi d\Omega^2_{p-2})\, .
\end{equation}

The {\em horospheric coordinates} used by Gubser, Klebanov and Polyakov 
\cite{gkp} are defined as

\begin{equation}
\left\{
\begin{array}{rcl}
X^0 &=& t/z\, ,\\
& & \\
X^a &=& x^a /z\, ,\\
& & \\
X^p - X^{p-1}& =& 1/z\, ,\\
\end{array}
\right.
\end{equation}

\noindent (where $a = 1,\ldots,p-2 $)

The metric reads

\begin{equation}
ds^2 = \frac{1}{z^2} (dt^2 - d\vec{x}_{p-2}^2 - dz^2)\, ,
\end{equation}

\noindent where $d\vec{x}_{p-2}^2$ is the Euclidean line element in 
$\mathbb{R}_{p-2}$.

The metric above enjoys a manifest $O(1,p-2)$ symmetry. Besides, it is
invariant under {\em dilatations} $x^{\m}\rightarrow \lambda x^{\m}$
($ x^{\m} = (t,\vec{x},z)$) and {\em inversions} $x^{\m}\rightarrow
\frac{x^{\m}}{x^2}$. Of course those transformations just convey an
action of $O(2,p-1)$ on the horospheres.

The coordinates used by Maldacena in \cite{malda} are $u \equiv
\frac{R^2}{z}$, $R^2$ being the square radius of the total space. In
order to allow for this, we have to multiply the previous line element
by $R^2$, getting in that way Maldacena's metric:

\begin{equation}
ds^2 = - \frac{R^2}{u^2} du^2 + \frac{u^2}{R^2} dx_{\parallel}^2\, ,
\end{equation}

\noindent (where $dx_{\parallel}^2$ stands for the ordinary Minkowski
 metric in $M_{p-1}$)
 
 Horospheric coordinates break down at $z = \infty$ (u = 0), (which we
 shall call the {\em horizon}); which in terms of the embedding is
 just $X^p = X^{p-1}$. In terms of the global static coordinates of
 (A.35), this equation has solution for a given $\t$ for all $\chi <
 \chi(\t)\equiv \sinh^{-1} |\tan \t |$.

This region can be easily parametrized, using $(X^0)^2 - \d_{ij}X^i
X^j = 1$ $(i = 1\ldots p-2)$ by $X^0 = \cosh z; X^i = n^i \sinh z$,
and the induced metric on the horizon is:

\begin{equation}
ds^2 = -dz^2 - \sinh^2 z d\Omega^2_{p-3}\, .
\end{equation}

In order to study the (conformal) boundary using Penrose's
construction, \cite{he} we perform the further change (from the
globally static form): 

\begin{equation}
\xi \equiv 2 \left( \tan^{-1} (e^{\chi}) -{\textstyle\frac{\pi}{4}}\right)\, , 
\end{equation}

\noindent and we get 

\begin{equation}
ds^2_{AdS} = \cosh^2 \chi \, ds^2_{ESU}\, ,
\end{equation}

\noindent where the metric of the Einstein Static Universe is
given by 

\begin{equation}
ds^2_{ESU} = d\t ^2 - d\xi^2 -\sin^2 \xi d\Omega^2_{p-2}\, ,
\end{equation}

\noindent (where $ - \infty \leq \t \leq \infty$ and 
$ 0 \leq \xi \leq \pi $)

$CAdS$ corresponds to the portion of ESU represented by $0 \leq \xi
\leq \pi/2$. 

The boundary of $AdS$ can be identified with the surface $z=0$, which in
terms of the embedding coordinates is equivalent to $X^p - X^{p-1} =
\infty $.  This sits in the ESU on $\xi = \pi/2$.

It is then plain that in this way of looking at things the boundary
itself is part of the horizon. Let us stress once more that this
latter concept is not invariant under conformal rescalings.



\end{document}